\documentclass[12pt,preprint]{aastex61}
\usepackage[utf8]{inputenc}
\usepackage{xcolor}
\usepackage{graphicx}
\usepackage{natbib}
\usepackage{amsmath}
\RequirePackage{lineno}
\usepackage[version=3]{mhchem}

\renewcommand{\thefootnote}{\fnsymbol{footnote}}

\begin{document}

\title{Effects of Mg/Si on Exoplanetary Refractory Oxygen Budget}

\author{Cayman T. Unterborn}
\altaffiliation{SESE Exploration Fellow}
\affiliation{School of Earth and Space Exploration, Arizona State University, Tempe, AZ 85287, USA}
\email{cayman.unterborn@asu.edu}

\author{Wendy R. Panero}
\affiliation{School of Earth Sciences, The Ohio State University, Columbus, OH, 43210}

\received{April 28, 2016}
\revised{April 4, 2017}

\begin{abstract}
Solar photospheric abundances of refractory elements mirror the Earth's to within $\sim$10 mol\% when normalized to the dominant terrestrial planet-forming elements Mg, Si and Fe. This allows for the adoption of Solar composition as an order-of-magnitude proxy for Earth's. It is not known, however, the degree to which this mirroring of stellar and terrestrial planet abundances holds true for other star-planet systems without determination of the composition of initial planetesimals via condensation sequence calculations and post condensation processes. We present the open-source Arbitrary Composition Condensation Sequence calculator (ArCCoS) to assess how the elemental composition of a parent star affects that of the planet-building material, including the extent of oxidation within the planetesimals. We demonstrate the utility of ArCCoS by showing how variations in the abundance of the stellar refractory elements Mg and Si affect the condensation of oxygen, a controlling factor in the relative proportions of planetary core and silicate mantle material. This, thereby, removes significant degeneracy in the interpretation of the structures of exoplanets as well as providing observational tests for the validity of this model.
\end{abstract}

\section{Introduction}
\label{sec:intro}
\indent The most abundant element in the Earth is a nominally volatile one: oxygen. Constituting up $\sim$50\% of the planet's total atoms \citep{McD03}, oxygen plays an important role at all scales within the Earth; controlling the bulk structure of the Earth, the speciation of minerals within the mantle and, indeed, its surface habitability. For exoplanets, however, the structure and composition of these planets is currently determined through inference from planetary mass and radius \citep[e.g.][]{Vale06, Sot07,Seag07,Fort07,Roge10,Wagn11,Zeng13, Lope14,Unte16}, with surface habitability limited to potential atmospheric observations \citep{Seag15}. Inverse models of terrestrial planets \citep[e.g. ][]{Dorn15} find these observables are sufficient only to constrain central core size and only where an atmosphere is assumed to be a negligible component of the planet's mass. Without this fully terrestrial assumption, mass-radius models cannot a-priori determine whether a planet is a super-Earth or a mini-Neptune, that is, whether a planet is a massive terrestrial planet or contains a significant gaseous envelope. This is due to the first-order trade off in the size of the planet's metallic core with the thickness of its atmosphere.  \\
\indent \citet{Dorn15} point out that there is considerable compositional and mineralogical degeneracy present in these models, that is mulitple planetary interior compositions are valid solutions given only the constraints of a planet's total mass and radius. They show that adopting a host star's abundance of the refractory planet-building elements (Mg, Si and Fe) as a proxy for planetary composition, reduces this compositional degeneracy. \citet{Unte16} expands upon this approach by demonstrating that the Sun's refractory composition is a sufficient proxy for reproducing the Earth's bulk structure and mineralogy, but only by adopting a liquid iron core, upper mantle structure and realistic light element budget for the core; all aspects lacking in the ``canonical'' mass-radius model of \citet{Zeng13}. These omissions cause the model to systematically overestimate the mass of Earth and ``Earth-like'' planets for a given radius. \citet{Unte16} further created a grid of benchmarked mass-radius models in order to quantitatively define ``Earth-like'' planets. The oxygen abundance of a planet in both models, however, was determined only as a consequence of the relative proportions of the core to mantle, rather than assumed to be the stellar value. As such, both models make broad assumptions on the bulk oxidation state of these exoplanets.\\
\indent The size of a planetary core is a direct consequence of a planet's oxidation state. Core formation results from early differentiation in the planetary formation process as a consequence of melting and immiscibility of metal and silicate, in which some fraction of the metal is not oxidized upon condensation. Therefore, to first-order, the fraction of free metal in the core is a function of the total oxygen abundance of a planet. The Sun contains $\sim$5 times more oxygen than the sum of the planet-building cations Mg, Si and Fe \citep{Aspl05}. Thus if one were to erroneously assume that, like the refractory elements, a planet's oxygen abundance is mirrored between planet and star, they would predict the Earth as a planet that is entirely oxidized, with no core present. This discrepancy is due to the dual nature of oxygen: refractory, condensing as silicates and oxides, and volatile, condensing as ice. As such, the assumptions of stellar abundance being indicative of planetary abundance that apply for the refractory elements as in \citet{Dorn15} and \citet{Unte16} cannot be assume for oxygen. We therefore have no direct or indirect way to measure the abundance of the most dominant terrestrial planet-building element.\\
\indent Equilibrium condensation models predict $\sim$23\% of Solar O entering into rocky phases \citep{Lodd03}. However, because oxygen behaves both as a refractory and volatile element, the fraction of oxygen condensing as a refractory, terrestrial-planet building component will necessarily be a function of the bulk composition of the nebula, in particular the relative proportions of dominant rock-forming elements (Ca, Al, Ti, Mg, Si, and Fe) relative to oxygen, but also the oxidation state of these elements within the solid. \\
\indent The existing condensation codes to calculate this speciation though are either not open-source \citep[e.g.][]{Ebel00,Lodd03} or proprietary (e.g. HSC Chemistry), hindering self-consistent comparison across studies and testing of thermodynamic databases across a range of compositions not necessarily relevant to the Solar System. As we seek to constrain the compositional and structural diversity of planetary systems outside of our own, we must understand how the variation in stellar compositions affect the composition of associated planets, and therefore their structure and mineralogy and thus the likelihood of it being ``Earth-like.'' We present then, the Arbitrary Composition Condensation Sequence Calculator (ArCCoS) with a flexible thermodynamic database. ArCCoS calculates the stable assemblages of a gas and solid in equilibrium to determine the relative proportions and stoichiometry of the refractory condensing phases from which terrestrial planets are built. \footnote[1]{The code and documentation is available for download at: \href{https://github.com/CaymanUnterborn/ArCCoS}{https://github.com/CaymanUnterborn/ArCCoS}} \\
\indent The chemical composition of the Earth is constrained from chondritic models, source material from the upper-mantle \citep{McD95, McD03, Javo10}, and geophysical constraints such as total mass of the planet, moment of inertia measurements, and seismic wave speeds. For this Earth composition, the dominant minerals by volume are the lower mantle minerals bridgmanite, perovskite structure (Mg,Fe)SiO$_{3}$ and ferropericlase, (Mg,Fe)O. The relative proportions of these two minerals is a function of the mantle's Mg/Si ratio. We highlight one potential utility of the ArCCoS code such by examining the effects of variable stellar Mg/Si on the percentage of condensed oxygen into the refractory solid phases and the subsequent changes expected in the bulk structure of the resulting planet.
\section{Methods}
\begin{deluxetable*}{llllllllll}
\tablecolumns{10}
\tablewidth{0pt}
\tablecaption{Gas species included in ArCCoS. All thermochemical data taken from \citet{NIST} unless noted. \label{Gasses}}
  \leavevmode
  \centering
   \epsfxsize=1cm
\tablehead{ \multicolumn{0}{c}{\phantom} }
\startdata 
AlS&CH$_{4}$&C$_{2}$K$_{2}$N$_{2}$&H$_{3}$N&F$_{2}$S&S$_{2}$&ClO&Cl$_{3}$OP&F$_{2}$O&MgF\\
AlF$_{2}$&CNO&C$_{2}$HF&K$_{2}$&PS&CCl$_{3}$F&ClTi&CaH$_{2}$O$_{2}$&F$_{2}$Ti&FN\\
CCl$_{2}$F$_{2}$&CH$_{2}$&Na$_{2}$&H$_{2}$MgO$_{2}$&F$_{3}$NO&F0S$_{2}$&ClH&CrO$_{2}$&TiF&FNO\\
AlF$_{2}$O&CN$_{2}$-trans&C$_{2}$H&MgN&O$_{3}$S&CHCl&Cl$_{2}$Co&HNO$_{2}$-trans&F$_{3}$OP&Al\\
AlClF&CNNa&C$_{2}$H$_{2}$&KO&P$_{2}$&CClF$_{3}$&ClP&Cl$_{2}$O&F$_{2}$P&Ar\\
AlHO-cis&CS&CF$_{2}$&NO$_{3}$&PO$_{2}$&CaCl&HS&CaCl$_{2}$&F$_{5}$P&C\\
AlClO&COS&C$_{2}$F$_{4}$&MgS&TiO$_{2}$&CHF&Cl$_{2}$K$_{2}$&HNa&F$_{3}$PS&Ca\\
CClN&CH$_{2}$O&C$_{2}$O&H$_{2}$Na$_{2}$O$_{2}$&ClS&CHClF$_{2}$&ClMg&Cl$_{5}$P&F$_{2}$N&Cl\\
AlHO$_{2}$&C$_{2}$&CF$_{3}$&NS&FNO$_{2}$&F$_{2}$S$_{2}$-cis&HO$_{2}$&CaF$_{2}$&HK&Co\\
AlN&C$_{2}$Cl$_{2}$&CF$_{4}$&NSi&TiO&F$_{2}$S$_{2}$-trans&HO&CaHO&HKO&Cr\\
AlO&C$_{2}$Cl$_{4}$&CF$_{4}$O&N$_{2}$&SiO&FHO$_{3}$S&HNaO&CaO&HMg&F\\
AlH&CP&CHNO&NO$_{2}$&O$_{2}$S&CClFO&Cl$_{2}$Na$_{2}$&C$_{5}$&F$_{4}$S&Fe\\
CCl$_{3}$&CHO&N$_{3}$&HSi&F$_{7}$H$_{7}$&S$_{4}$&ClF$_{3}$&FH&FP&H\\
CCl$_{2}$O&CHP&NaO&H$_{2}$&H$_{2}$K$_{2}$O$_{2}$&S$_{3}$&ClF$_{5}$&CrO$_{3}$&FPS&He\\
AlCl$_{2}$&CN$_{2}$-cis&C$_{2}$HCl&K$_{2}$O$_{4}$S&O$_{6}$P$_{4}$&CF$_{8}$S&Cl$_{2}$&HNO$_{2}$-cis&F$_{3}$N&K\\
Al$_{2}$O&CH$_{3}$F&C$_{2}$N&H$_{2}$S&FS&SSi&ClNa&Cl$_{3}$P&F$_{2}$Na$_{2}$&Mg\\
AlF&CO$_{2}$&C$_{2}$F$_{2}$&NO&O$_{2}$Si&CH$_{2}$Cl$_{2}$&Cl$_{2}$Mg&C$_{4}$N$_{2}$&F$_{4}$N$_{2}$&Mn\\
CCl$_{2}$&CH$_{2}$ClF&PO&H$_{2}$N&NiS&ClS$_{2}$&ClHO&CrO&F$_{2}$&N\\
Al$_{2}$O$_{2}$&CH$_{3}$Cl&C$_{2}$N$_{2}$&H$_{2}$O$_{4}$S&Cl$_{2}$S&P$_{4}$S$_{3}$&ClNO$_{2}$&Cl$_{3}$PS&F$_{2}$N$_{2}$-trans&Na\\
CCl&CH$_{3}$&C$_{2}$N$_{2}$Na$_{2}$&H$_{2}$O&ClF$_{5}$S&P$_{4}$&ClNO&Cl$_{4}$Co$_{2}$&F$_{2}$N$_{2}$-cis&Ne\\
AlHO-trans&CS$_{2}$&CF$_{2}$O&NP&O$_{2}$&CrN&PH&CaF&F$_{6}$S&Ni\\
CClO&CH$_{2}$F$_{2}$&C$_{3}$&H$_{2}$N$_{2}$&FeO&ClFO$_{2}$S&ClK&CoF$_{2}$&F$_{2}$K$_{2}$&O\\
AlO$_{2}$&CKN&C$_{2}$H$_{4}$O&H$_{3}$P&F$_{3}$S&S$_{8}$&ClOTi&Cl$_{3}$Co&F$_{2}$OS&P\\
Al$_{2}$&CN&C$_{2}$H$_{4}$&H$_{4}$N$_{2}$&Cl$_{2}$O$_{2}$S&AlFO&ClO$_{2}$&Cl$_{2}$Ti&F$_{2}$O$_{2}$S&S\\
CCl$_{4}$&CHN&N$_{2}$O$_{5}$&C$_{4}$&F$_{6}$H$_{6}$&S$_{5}$&ClF$_{2}$OP&FHO&FO$_{2}$&Si\\
AlCl&CO&C$_{2}$F$_{6}$&MgO&O$_{3}$&CHCl$_{2}$F&Cl$_{2}$FOP&HNO$_{3}$&F$_{3}$P&Ti\\
CF&CHF$_{3}$&Na$_{2}$O$_{4}$S&C$_{3}$O$_{2}$&F$_{5}$H$_{5}$&S$_{6}$&ClFO$_{3}$&FK&FOTi&Cl$_{2}$Cr$^{\dagger}$\\
CFN&CHFO&N$_{2}$O$_{4}$&H$_{2}$P&F$_{4}$H$_{4}$&S$_{7}$&ClF&HNO&FNa&Cl$_{2}$CrO$_{2}$$^{\dagger}$\\
CFO&CHCl$_{3}$&N$_{2}$O$_{3}$&OS&F$_{3}$H$_{3}$&C$_{2}$H$_{4}$O&ClCo&HN&FNO$_{3}$&CrS$^{\dagger}$\\
C$_{2}$Cl$_{6}$&CH&N$_{2}$O&OS$_{2}$&F$_{2}$H$_{2}$&P$_{4}$O$_{10}$&CaS&HMgO&NiO$^{\dagger}$&F$_{2}$Mn$^{\dagger}$\\
Cl$_{2}$Mn$^{\dagger}$&TiS$^{\dagger}$&CoO$^{\ddagger}$&MnO$^{\ddagger}$&&&&&\\
\enddata
\tablerefs{\scriptsize $^{\dagger}$: \citet{Knac91}; $^{\ddagger}$: \citet{Pedl83}}
\end{deluxetable*}
\begin{deluxetable*}{llll|lll}
\tablecolumns{7}
\tablewidth{0pt}
\tablecaption{List of solid species included in ArCCoS including references. \label{Solids}}
\tablehead{\multicolumn{4}{c}{Misc. Solids} & \multicolumn{3}{c}{\citet{NIST}}}
  \leavevmode
  \centering
   \epsfxsize=1cm
\startdata 
\AA kermanite&Ca$_{2}$MgSi$_{2}$O$_{7}$$^{\dagger}$&Grossite&CaAl$_{4}$O$_{7}$$^{\mathsection}$&Al&FNa&Mg$_{2}$Si\\
Albite&NaAlSi$_{3}$O$_{8}$$^{\dagger}$&Grossular&Ca$_{3}$Al$_{2}$Si$_{3}$O$_{12}$$^{\dagger}$&AlN&F$_{2}$Fe&Mn\\
Almandine&Fe$_{3}$Al$_{2}$Si$_{3}$O$_{2}$$^{\dagger}$&Hibonite&CaAl$_{2}$O$_{19}$$^{\ddagger}$&Al$_{2}$S$_{3}$&F$_{2}$Mg&N$_{4}$Si$_{3}$\\
Andalusite &Al$_{2}$SiO$_{5}$$^{\dagger}$&Ilmenite&FeTiO$_{3}$$^{\dagger}$&Al$_{6}$Si$_{2}$O$_{13}$&Fe&Na\\
Anorthite&CaAl$_{2}$Si$_{2}$O$_{8}$$^{\dagger}$&Jadeite&NaAlSi$_{2}$O$_{6}$$^{\dagger}$&C&FeH$_{2}$O$_{2}$&NaO$_{2}$\\
Anthophyllite &Mg$_{7}$Si$_{8}$O$_{22}$(OH)$_{2}$$^{\dagger}$&Lime&CaO$^{\dagger}$&C$_{2}$Mg&FeH$_{3}$O$_{3}$&Na$_{2}$O$_{3}$Si\\
Brucite &Mg(OH)$_{2}$$^{\dagger}$&Magnesite&MgCO$_{3}$$^{\dagger}$&C$_{3}$Al$_{4}$&FeO&Na$_{2}$SO$_{4}$-$\delta$\\
Ca-aluminate&CaAl$_{2}$O$_{4}$$^{\mathsection}$&Magnetite&Fe$_{3}$O$_{4}$$^{\dagger}$&C$_{3}$Mg$_{2}$&FeO$_{4}$S&Na$_{2}$SO$_{4}$-III\\
Calcite&CaCO$_{3}$$^{\dagger}$&Merwinite&Ca$_{3}$MgSi$_{2}$O$_{8}$$^{\dagger}$&Ca-$\alpha$&FeS$_{2}$-I&Ni\\
Clinoenstatite&MgSiO$_{3}$$^{\dagger}$&Monticellite&CaMgSiO$_{4}$$^{\dagger}$&Ca-$\beta$&FeS$_{2}$-II&P-I\\
Corderite&Mg$_{2}$Al$_{4}$Si$_{5}$O$_{18}$$^{\dagger}$&Periclase&MgO$^{\dagger}$&CaCl$_{2}$&Fe$_{2}$O$_{2}$S$_{3}$&S\\
Corundum&Al$_{2}$O$_{3}$$^{\dagger}$&Perovskite&CaTiO$_{3}$$^{\sharp}$&CaH$_{2}$O$_{2}$&HK&S$_{2}$Si\\
Cristobalite&SiO$_{2}$$^{\dagger}$&Pyrope&Mg$_{3}$Al$_{2}$Si$_{3}$O$_{12}$$^{\dagger}$&CaS&HNa&SiC-$\alpha$\\
Diopside&CaMgSi$_{2}$O$_{6}$$^{\dagger}$&Quartz-$\alpha$&SiO$_{2}$$^{\dagger}$&ClK&H$_{2}$Mg&SiC-$\beta$\\
Dolomite&CaMg(CO$_{3}$)$_{2}$$^{\dagger}$&Sanidine&KAlSi$_{3}$O$_{8}$$^{\dagger}$&ClNa&K&Ti-$\alpha$\\
Enstatite&MgSiO$_{3}$$^{\dagger}$&Sillimanite&Al$_{2}$SiO$_{5}$$^{\dagger}$&Cl$_{2}$Fe&K$_{2}$O&Ti-$\beta$\\
Fayalite&Fe$_{2}$SiO$_{4}$$^{\dagger}$&Sphene&CaTiSiO$_{5}$$^{\dagger}$&Cl$_{2}$Mg&K$_{2}$O$_{3}$Si&TiH$_{2}$\\
Ferrosilite&FeSiO$_{3}$$^{\dagger}$&Spinel&MgAl$_{2}$O$_{4}$$^{\dagger}$&Co&Mg&TiO-$\alpha$\\
Forsterite&Mg$_{2}$SiO$_{4}$$^{\dagger}$&Talc&Mg$_{3}$Si$_{4}$(OH)$_{2}$$^{\dagger}$&CoO&MgO$_{4}$S&TiO-$\beta$\\
Gehlenite&Ca$_{2}$Al$_{2}$SiO$_{7}$$^{\dagger}$&&&FK&MgS&Ti$_{3}$O$_{5}$-$\alpha$ \\
Wollastonite&CaSiO$_{3}$$^{\dagger}$&&&Ti$_{3}$O$_{5}$-$\beta$&\\
\enddata
\tablerefs{\scriptsize $^{\dagger}$: \citet{Berm88}; $^{\mathsection}$: C$_{p}$ from \citet{Berm85}; 298 K data from \citet{Berm83}; $^{\ddagger}$: \citet{Berm83}; $^{\sharp}$: \citet{Robi78}}
\end{deluxetable*}
\indent Conservation of mass and the law of mass action determine the condensation temperature of a solid in equilibrium with nebular gas. The governing equations are functions of the total pressure of the system ($P_{tot}$), the number of moles of each element in the system and the distribution of these elements between each species in equilibrium. Mass balance is the sum of the number density, $n$ in mol L$^{-1}$, of element or compound, $X$, in each gas phase, $i$, and solid phase, $j$: 
\begin{equation}
\label{mass_balance}
N_{X} = \sum_{i}\nu_{i,X}n_{i} + \sum_{j} \nu_{j,X}n_{j}
\end{equation}
where $\nu$ is the stoichiometric coefficient of compound $i$ or $j$ in the individual phases. The distribution of an element $X$ between the gas and solid phases can then calculated according to the law of mass action via the equilibrium constant, $K_{i,j}$:
\begin{equation}
ln(K_{i,j}) = ln\left (\frac{n_{i,j}}{\prod_{X}^{}(n_{X})^{\nu_{(i,j),X}}}* RT^{\left(\sum_{X}^{}\nu_{(i,j),X}\right)-1}\right) = \frac{-\Delta G_{r}}{RT}
\label{eqconst}
\end{equation}
\noindent where $n$ is the number density of a gas/solid/element $i$, $j$ or $X$, $R$ is the gas constant, $T$ is the temperature (in K) and $\Delta G_{r}$ is the Gibbs free energy of the reaction from the elements as defined by:
\begin{equation}
\Delta G_{r} = \Delta G_{i,j} - \Delta G_{elements}
\end{equation}
\indent Assuming each gas phase is ideal and at constant volume, $N_{X}$ is proportional to the partial pressure of element $X$ in the system via:
\begin{equation}
\label{par_pres}
N_{X} = \frac{a(X)}{\sum_{i}a_{i}+\sum_{j}a_{j}}\ast\frac{P_{tot}}{RT}
\end{equation}
where $a$ is the number of moles of element $X$ from the input Solar model or the number of moles of gas/solid, $i,j$ in the system. We simplify equation \ref{par_pres}, by only considering the dominant species by mole in the system: H, H$_{2}$ and the noble gasses (He, Ne, Ar):
\begin{equation}
N_{X} = \frac{a(X)}{a(\text{H+H$_{2}$+noble gasses})}\ast\frac{P_{tot}}{RT}
\label{massbal1}
\end{equation}
 The temperature-dependent distribution of $a$(H) and $a$(H$_{2}$) are determined from their equilibrium coefficients taken from the JANAF tables \citep{NIST}. At present, ArCCoS determines the speciation between $\sim$400 gas species, and 23 elements with $\sim$110 potential solid condensates (Table \ref{Gasses}). The ArCCoS database includes 23 elements: H, He, C, N, O, F, Ne, Na, Mg, Al, Si, P, S, Cl, Ar, K, Ca, Ti, Cr, Mn, Fe, Co, and Ni. Of these, we treat Si, Mg, Ca, Al, Ti, Ni, and Fe as the refractory elements.\\
\indent All reactions are considered to occur at a given $P$ and $T$ from the reference elements with the exception of H, N, O, F and Cl where the molecular form (e.g. O$_{2}$) is chosen as the reference form. Data for these reference phases for enthalpy and entropy were adopted from the JANAF tables \citep{NIST} with linear interpolation in $T$. The Gibbs free energy of the reactant elements is therefore:
\begin{equation}
\Delta G_{elements} = \sum_{X}^{}\nu_{X} \left [ H(T) - T \ast S(T) \right ]_{X}
\end{equation}
where $H(T)$ is the enthalpy of formation at temperature $T$, $S$ is the entropy and $\nu_{X}$ is the stoichiometric constant of element $X$ in the reaction.
\subsection{Gasses}
Above 2000 K, no solids are stable and only gas-phase equilibria need be considered. For example: the gas reaction of water from its constituent elements in their reference states: 
\begin{equation}
\ce{H_{2}(g) + \frac{1}{2}O_{2}(g) \leftrightarrow H_{2}O(g)}
\end{equation}
At a given $T$, the free energies of each species is known and their equilibrium coefficients can be written in terms of their partial pressures ($p_{X}$):
\begin{equation}
K_{H_{2}O} = \frac{p_{H_{2}O}}{p_{H_{2}}p_{O_{2}}^{0.5}}
\end{equation}
The number density of gaseous water ($n_{H_{2}O}$) is then taken from the ideal gas law:
\begin{equation}
n_{H_{2}O} = \frac{p_{H_{2}O}}{RT} = n_{H_{2}}n_{O_{2}}^{0.5}K_{H_{2}O}(RT)^{0.5}
\end{equation}
where $n_{H_{2}}$ and $n_{O_{2}}$ are the number densities of molecular hydrogen and oxygen, respectively. Similar equations are written for each gaseous species. For example, the mass balance equation for oxygen is written:
\begin{equation}
N_{O} = n_{H_{2}O} + n_{CO} + 2n_{CO_{2}} + \cdots
\end{equation}
which in turn can be written in terms of the concentrations component reference elements:
\begin{align}
N_{O} &= n_{H_{2}}n_{O_{2}}^{0.5}K_{H_{2}O}(RT)^{0.5} + n_{C}n_{O}^{0.5}K_{CO_{2}}(RT)^{0.5} \nonumber\\
&+ 2 n_{C}n_{O_{2}}K_{CO_{2}}(RT) + \cdots
\label{massbal2}
\end{align}
Combining equations \ref{massbal1} and \ref{massbal2} for each of the 23 elements in the system solved in ArCCoS represents a system of 23 nonlinear equations.\\
\indent 
For the gas phases (Table \ref{Gasses}), the thermodynamic data are from \citet{NIST}, \citet{Knac91} and \citet{Pedl83}. At temperatures within the range reported for a given species in \citet{NIST} and \citet{Pedl83}, $\Delta G_{gas}$ as a function of temperature is determined by linear interpolation. For those species taken from \citet{Knac91}, we follow their methodology for determining the $\Delta G_{gas}$ for an individual gas phase.
\subsection{Solid Condensation}
Condensation of solids from the gas phase occurs when the partial pressure of the component elements exceeds the equilibrium constant and thus the relationship:
\begin{equation}
K_{j} - \prod_{X}^{}(n_{X}RT)^{\nu_{X}} = 0
\label{condense}
\end{equation} 
is achieved, here $\nu_{X}$ is the stoichiometric constant for each constituent element. For example, the first solid to condense for a system of Solar composition at $P_{tot}$ = 10$^{-3}$ bar is corundum at 1770 K when $P_{Al_{2}O_{3}} = K_{Al_{2}O_{3}}$. For corundum equation \ref{condense} is then written as:
\begin{equation}
K_{Al_{2}O_{3}} - n_{Al}^{2}n_{O_{2}}^{1.5}RT^{3.5} = 0
\label{condense_cor}
\end{equation} 
\begin{deluxetable}{lcccc}
\tablecolumns{5}
\tablewidth{0pt}
\tablecaption{Comparison of appearance (in) and disappearance (out) temperatures (in K) of solid phases between this study and the model of \citet{Ebel00}. Those solids with no reported disappearance temperature remain stable at the termination of the calculation. All calculations are run at $P_{tot}$ = 10$^{-3}$ bar. \label{Compare}}
\tablehead{\colhead{} & \multicolumn{2}{l}{\citet{Ebel00}} & \multicolumn{2}{c}{This study} \\
\colhead{Solid} & \colhead{In} &\colhead{Out} &\colhead{In} &\colhead{Out}}
\startdata 
Corundum&1770&1726&1771&1731\\
Hibonite&1728&1686&1735&1699\\
Grossite&1698&1594&1712&1592\\
Perovskite&1680&1458&1681&1396\\
CaAl$_{2}$O$_{4}$&1624&1568&1623&1557\\
Melilite&1580&1434&1575&1456\\
(Gehlenite)&&&&\\
Grossite&1568&1502&1558&1486\\
Hibonite&1502&1488&&\\
Spinel&1488&1400&1487&1463\\
Melliilite&&&1457&1447\\
(\AA kermanite)&&&&\\
Fe&1462& &1453&\\
Clinopyroxene&1458& &1449&\\
Olivine&1444& &1446&\\
Plagioclase&1406&1318&1466&\\
Ti$_{3}$O$_{5}$&1368&1242&1397&1214\\
Ni&&&1382&\\
Orthopyroxene&1366&&1372&\\
Co&&&1272&\\
Corderite&1330&&&\\
Cr-Spinel&1230&&&\\
TiO$_{2}$&&&1215&\\
\enddata
\end{deluxetable}
The temperature at which the constraint in equation \ref{condense} is met is considered the initial or ``appearance'' condensation temperature. Once corundum begins to condense, some fraction, $n_{Al_{2}O_{3}}$, of the solid is now in equilibrium with the gas phase and each elemental number density in equation \ref{mass_balance} must adjust to this new constraint. As gas chemistry changes upon cooling, many of the first condensates become unstable, and rereact with the gas to form new phases. A solid is considered to be removed from the system when the solid's number density, $n_{j}$, falls below $1\ast10^{-10}$ mol per mol of atoms in the system following \citet{Shar95}. For corundum, this occurs at 1731 K (Table \ref{Compare}), with hibonite (CaAl$_{2}$O$_{19}$) becoming the new stable host of Al. This temperature is considered the solid's ``disappearance'' temperature. Its constituent element abundances are returned to the gas phase and the calculation is repeated. \\
\indent Gibbs free energy values of solid phases are either linearly interpolated when $\Delta G_{solid}$ is directly available or derived from reported specific heat functions (Table \ref{Solids}). In order to self-consistently calculate $\Delta G_{solid}$ from specific heat data we begin with the definition of $\Delta G_{solid}$:
\begin{equation}
\Delta G_{solid} = \Delta H(P,T) - T \ast S(P,T)
\end{equation}
where 
\begin{align}
\Delta H(P,T) &= \Delta H(P_{r},T_{r}) + \int_{T_{r}}^{T}C_{p}(T)dT \nonumber\\
&+ \int_{P_{r}}^{P} \left \{V(P_{r},T_{r})- T \left(\frac{\partial V}{\partial T} \right)_{P} \right \}dP 
\end{align}
and
\begin{equation}
S(P,T) = S(P_{r},T_{r}) + \int_{T_{r}}^{T}\frac{C_{p}(T)}{T}dT + \int_{P_{r}}^{P} \left(\frac{\partial V}{\partial T} \right)_{P} dP 
\end{equation}
where $\Delta H(P,T)$ and $S(P,T)$ are the enthalpy of formation from the elements and third law entropy at $P$ and $T$, $\Delta H(P_{r},T_{r})$ and $S(P_{r},T_{r})$ are the same values at reference $P$ and $T$ (1 bar, 298.15 K) and $V$ is the molar volume. The ambient pressures in these calculations are small ($P_{tot}$$\sim10^{-3}$ bar); we therefore ignore the volume integrals in our calculations of $\Delta G_{solid}$. As an exploratory study, we omit any formulation of solid solutions and consider only those condensates composed of pure, end-member phases.  The inclusion of solid solution models increases the complexity of the ArCCoS code and will be included in future updates and studies. \\
\subsection{Algorithm}
At each temperature, this system is solved using the \texttt{\bf scipy} root finding package with a least squares method using a modified Levenberg-Marquardt algorithm as implemented in MINPACK1 \citep{More80}. Models are run beginning at 2500 K, where only monoatomic gasses are present in the nebula and a solution can be easily found. ArCCoS requires an initial guess in order to begin solving equation \ref{mass_balance}. For the initial temperature calculation this is the concentration calculated in equation \ref{massbal1}, with subsequent initial guesses being the solution from the previous temperature step. Equilibrium is assumed when the sum of the least-squares difference in the mass-balance (Equation \ref{mass_balance}) for all elements is less than 10$^{-15}$ mol. Once a solution is found, the temperature is lowered by 2 K and the calculation repeated at the new temperature. ArCCoS continues solving equation \ref{mass_balance} until 100\% of each refractory element is condensed (Si, Mg, Fe, Ca, Al, Ni and Ti). The total percentage of oxygen in the refractory phases, \%RO, is then: 
\begin{equation}
\%RO= \frac{\sum_{j}^{}\nu_{j,O}n_{j}}{N_{O}}
\end{equation}
\subsection{Model Benchmark}
\begin{deluxetable}{lccc}
\tablecolumns{4}
\tablewidth{8cm}
\tablecaption{Stellar Abundances adopted in ArCCoS. Abundances are normalized such that $\log$(N$_{H}$) = 12.0 \label{Abundances}
}
\tablehead{\colhead{} & \colhead{Anders \&} & \colhead{Lodders} & \colhead{Asplund} \\
\colhead{Element} & \colhead{Grevesse (1989)} & \colhead{(2003)} & \colhead{et al. (2005)}}
\startdata
H&12.0&12.0&12.0\\
He&10.99$\pm$0.035&10.899$\pm$0.01&10.93$\pm$0.01\\
C&8.56$\pm$0.04&8.39$\pm$0.04&8.39$\pm$0.05\\
N&8.05$\pm$0.04&7.83$\pm$0.11&8.39$\pm$0.06\\
O&8.93$\pm$0.035&8.69$\pm$0.05&8.66$\pm$0.05\\
Ne&8.09$\pm$0.10&7.87$\pm$0.10&7.84$\pm$0.06\\
Mg&7.59$\pm$0.05&7.55$\pm$0.02&7.53$\pm$0.09\\
Al&6.48$\pm$0.07&6.46$\pm$0.02&6.37$\pm$0.06\\
Si&7.55$\pm$0.05&7.54$\pm$0.02&7.51$\pm$0.04\\
Fe&7.51$\pm$0.03&7.47$\pm$0.03&7.45$\pm$0.05\\
Ca&6.34$\pm$0.02&6.34$\pm$0.03&6.31$\pm$0.04\\
Ti&4.93$\pm$0.02&4.92$\pm$0.03&4.90$\pm$0.06\\
F&4.48$\pm$0.30&4.46$\pm$0.06&4.56$\pm$0.30\\
Cl&5.27$\pm$0.30&5.26$\pm$0.06&5.50$\pm$0.30\\
S&7.27$\pm$0.06&7.19$\pm$0.04&7.14$\pm$0.05\\
Na&6.31$\pm$0.03&6.30$\pm$0.03&6.17$\pm$0.04\\
Ar&6.56$\pm$0.10&6.55$\pm$0.08&6.18$\pm$0.08\\
Cr&5.68$\pm$0.03&5.65$\pm$0.05&5.64$\pm$0.10\\
Ni&6.25$\pm$0.04&6.22$\pm$0.03&6.23$\pm$0.04\\
P&5.57$\pm$0.04&5.46$\pm$0.04&5.36$\pm$0.04\\
K&5.13$\pm$0.13&5.11$\pm$0.05&5.08$\pm$0.07\\
Co&4.91$\pm$0.04&4.91$\pm$0.03&4.92$\pm$0.08\\
Mn&5.53$\pm$0.03&5.50$\pm$0.03&5.39$\pm$0.03\\
\enddata
\end{deluxetable}
\indent \citet[][hereafter EG00]{Ebel00} is the most similar model for benchmarking ArCCoS as we adopt the same input thermodynamic database and similar computational approach. For a gas at 10$^{-3}$ bar of the Solar composition reported in EG00 (Table \ref{Abundances}), our model predicts the same condensing solids with the exception of cordierite and Cr-spinel (a Cr rich solid solution of Mg and Cr spinel) at 1330 and 1230 K respectively (Table \ref{Compare}). Furthermore, our calculated appearance and disappearance are within $\sim$15 K. These discrepancies are likely a consequence of our non-incorporation of solid solutions in this model.\\
\indent Adopting the Solar abundances of \citet{Lodd03}, we calculate values similar to their reported appearance, 50\% condensation temperatures, the temperature at which 50\% of the element is condensed, and \%RO (Table \ref{Loddcomp}), with any difference being due to either the same solid-solution details discussed above or differences in the adopted input thermodynamic database, which \citet{Lodd03} does not report.
\begin{deluxetable}{lcc|cc}
\tablecolumns{5}
\tablecaption{Comparison of the initial condensation temperature and 50$\%$ condensation temperature for the major terrestrial planet-building elements between this work and \citet{Lodd03}. Calculations were performed using the input Solar abundances of \citet[][Table \ref{Abundances}]{Lodd03} and at $P_{tot}$=10$^{-4}$ bar \label{Loddcomp}}
\tablehead{\colhead{}&\multicolumn{2}{c}{T$_{appearance}$ (K)}&\multicolumn{2}{c}{T$_{50\%}$ (K)}\\
\colhead{} & \colhead{This} & \colhead{Lodders} & \colhead{This}&\colhead{Lodders}\\ 
\colhead{Element} & \colhead{Study} & \colhead{(2003)} & \colhead{Study}&\colhead{(2003)} }
\startdata
Al&1665&1677&1643&1653 \\
Ca&1609&1659&1508&1517\\
Ti&1583&1593&1569&1582\\
Mg&1397&1387&1336&1336\\
Si&1477&1529&1318&1310\\
Fe&1356&1357&1329&1310\\
\hline \\
Avg. Diff.&\multicolumn{2}{c}{22.5 K}&\multicolumn{2}{c}{9.8 K}\\
\enddata
\end{deluxetable}
\section{Results}
\begin{figure}
\begin{tabular}{ccc}
\includegraphics[width=.33\linewidth]{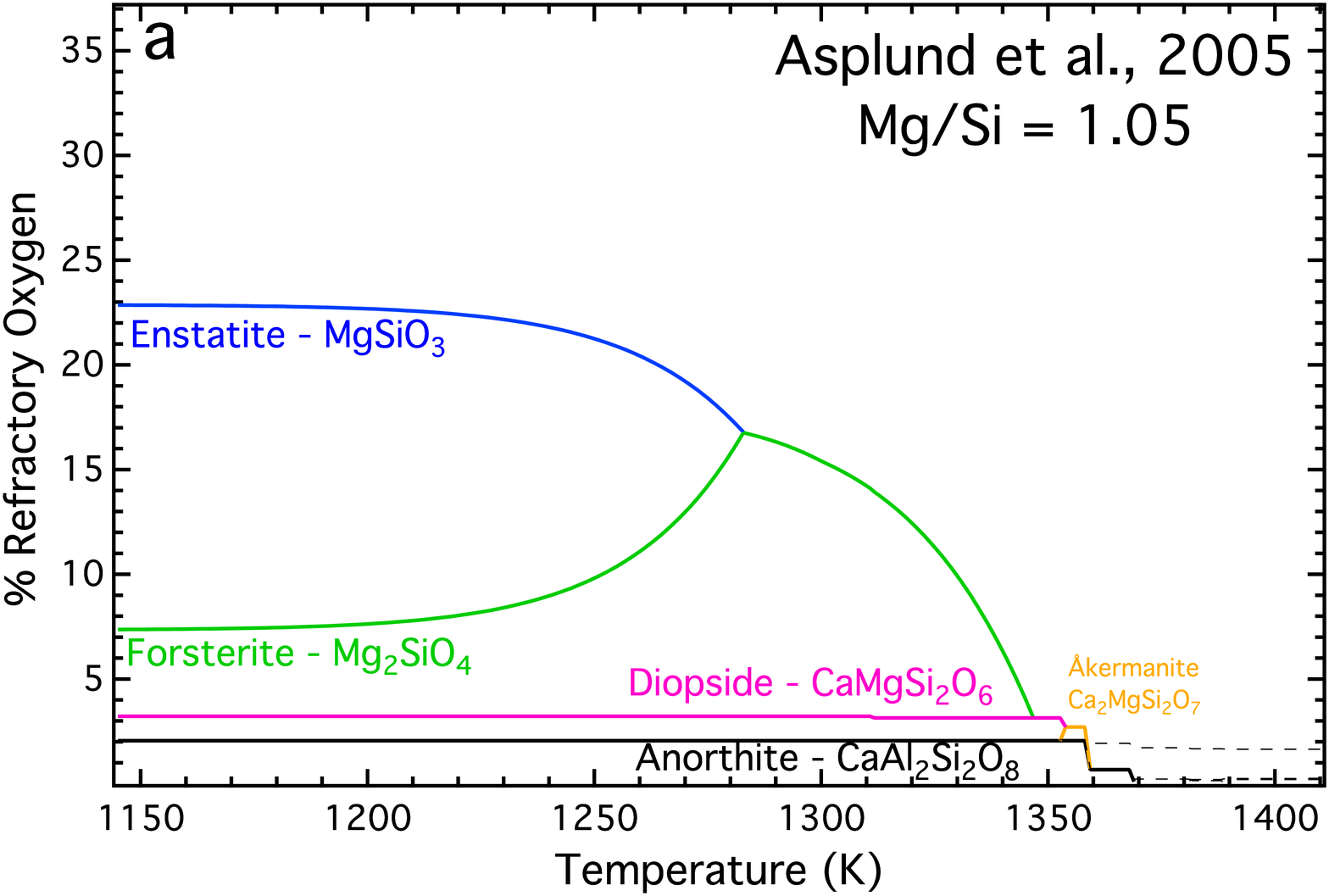}& \includegraphics[width=.33\linewidth]{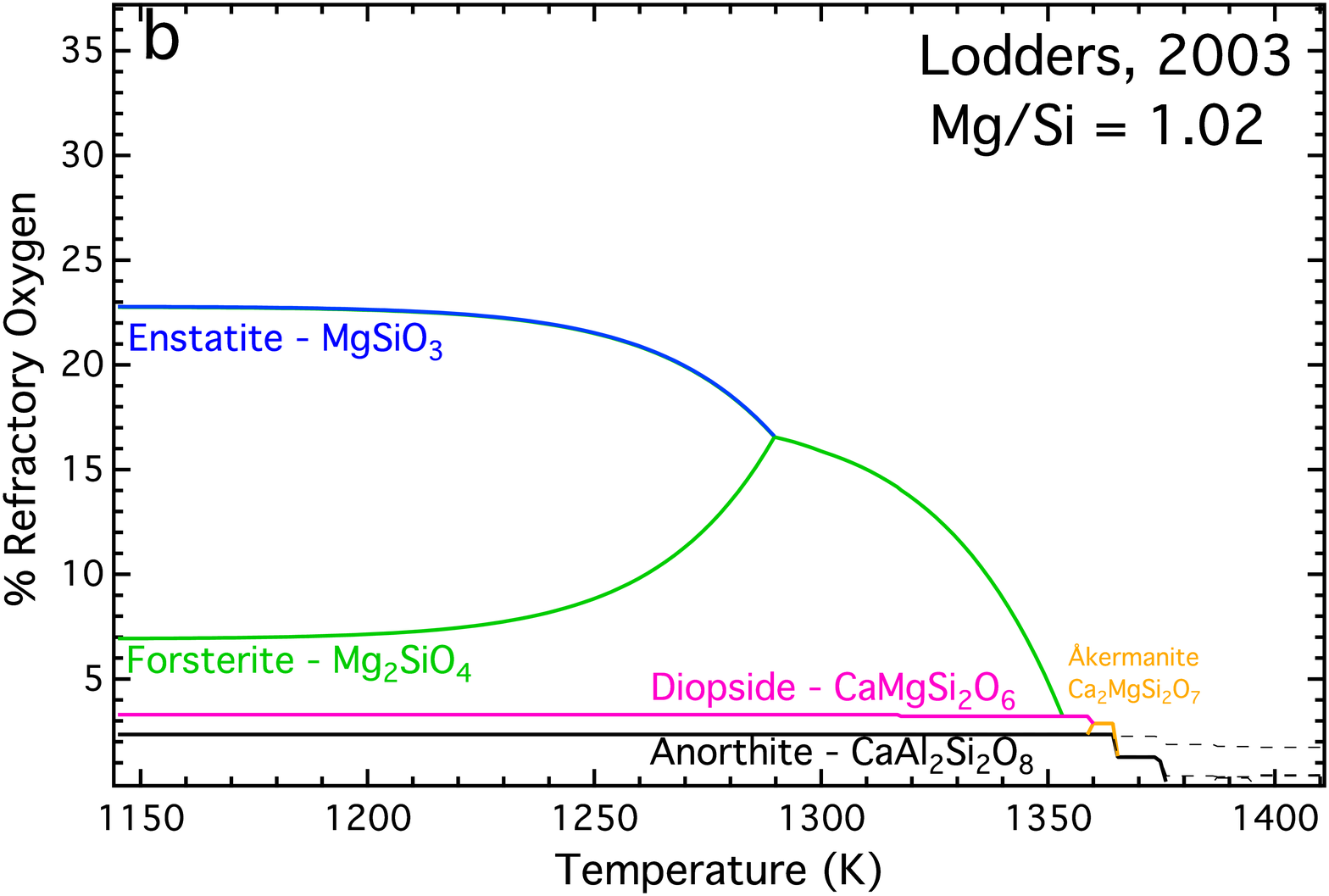}&\includegraphics[width=.33\linewidth]{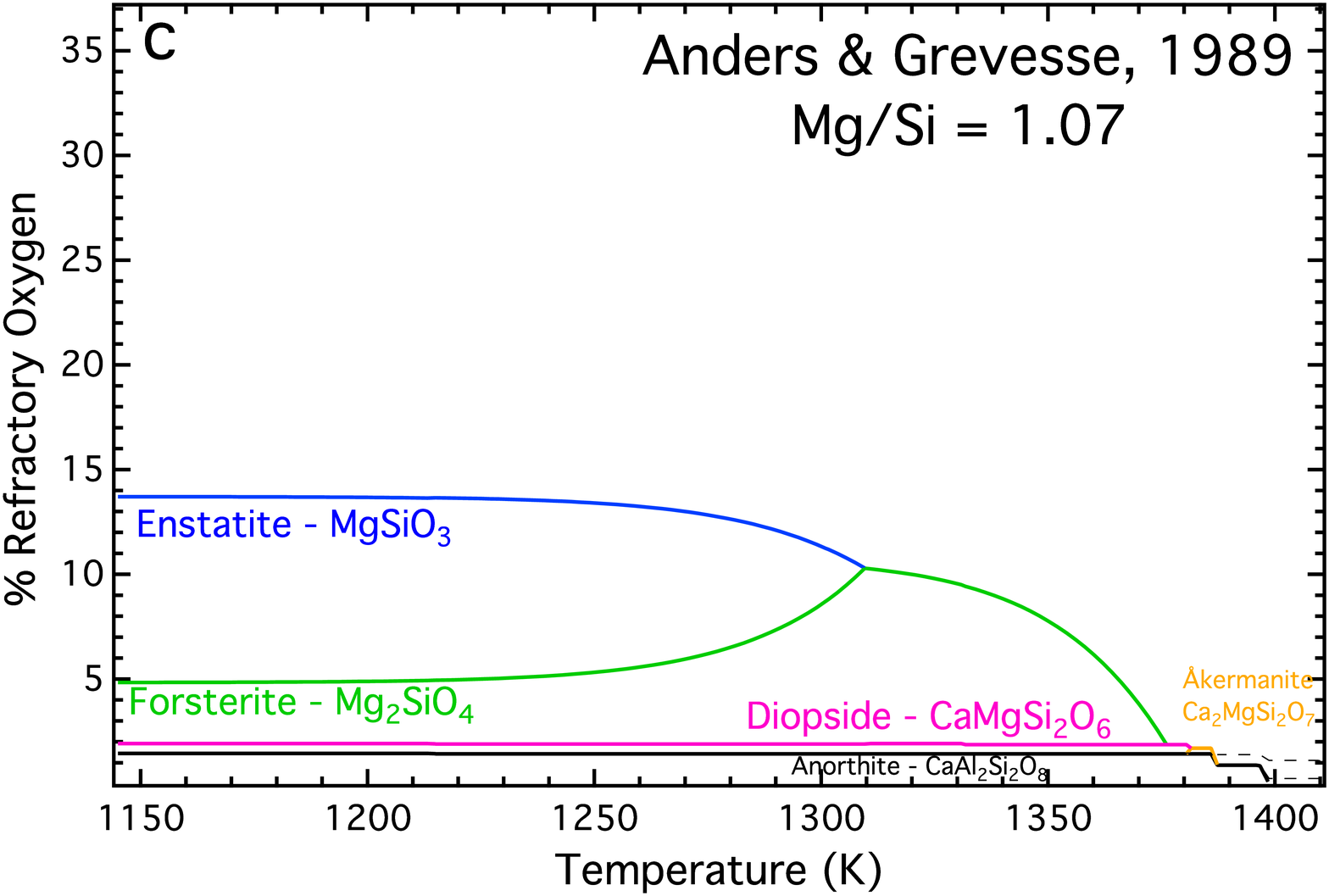}
\end{tabular}
\caption{O phase diagrams for condensation sequence calculations adopting the Solar composition of \citet{Aspl05,Lodd03} and \citet{Ande89} as inputs. All figures are on the same scale for comparison. Higher temperature condensed solids are shown as dashes.}
\label{SolarO}
\end{figure}

\begin{figure*}
  \leavevmode
  \centering
   \epsfxsize=8cm
\begin{tabular}{cc}
\includegraphics[width=8cm]{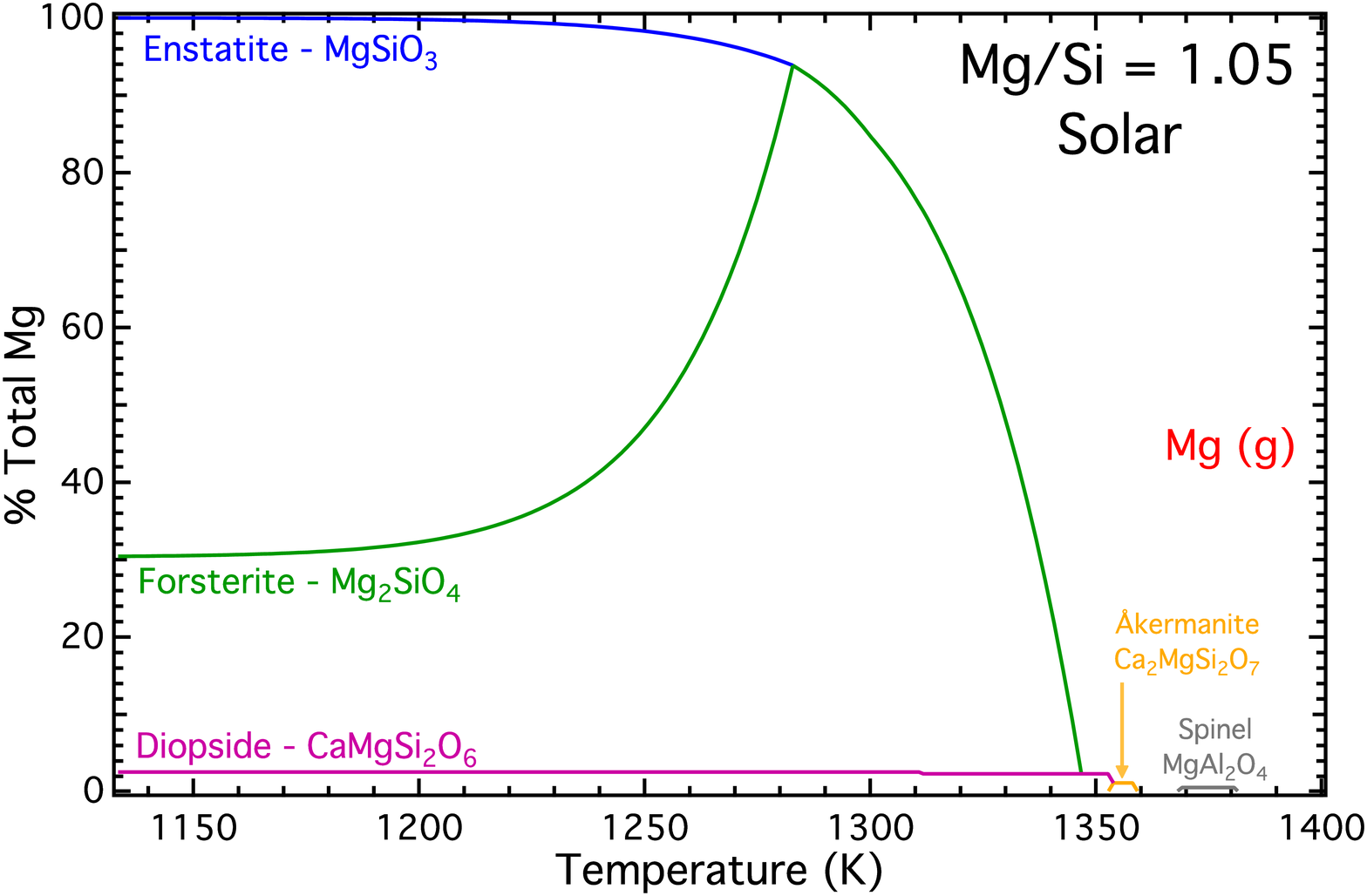} & \includegraphics[width=8cm]{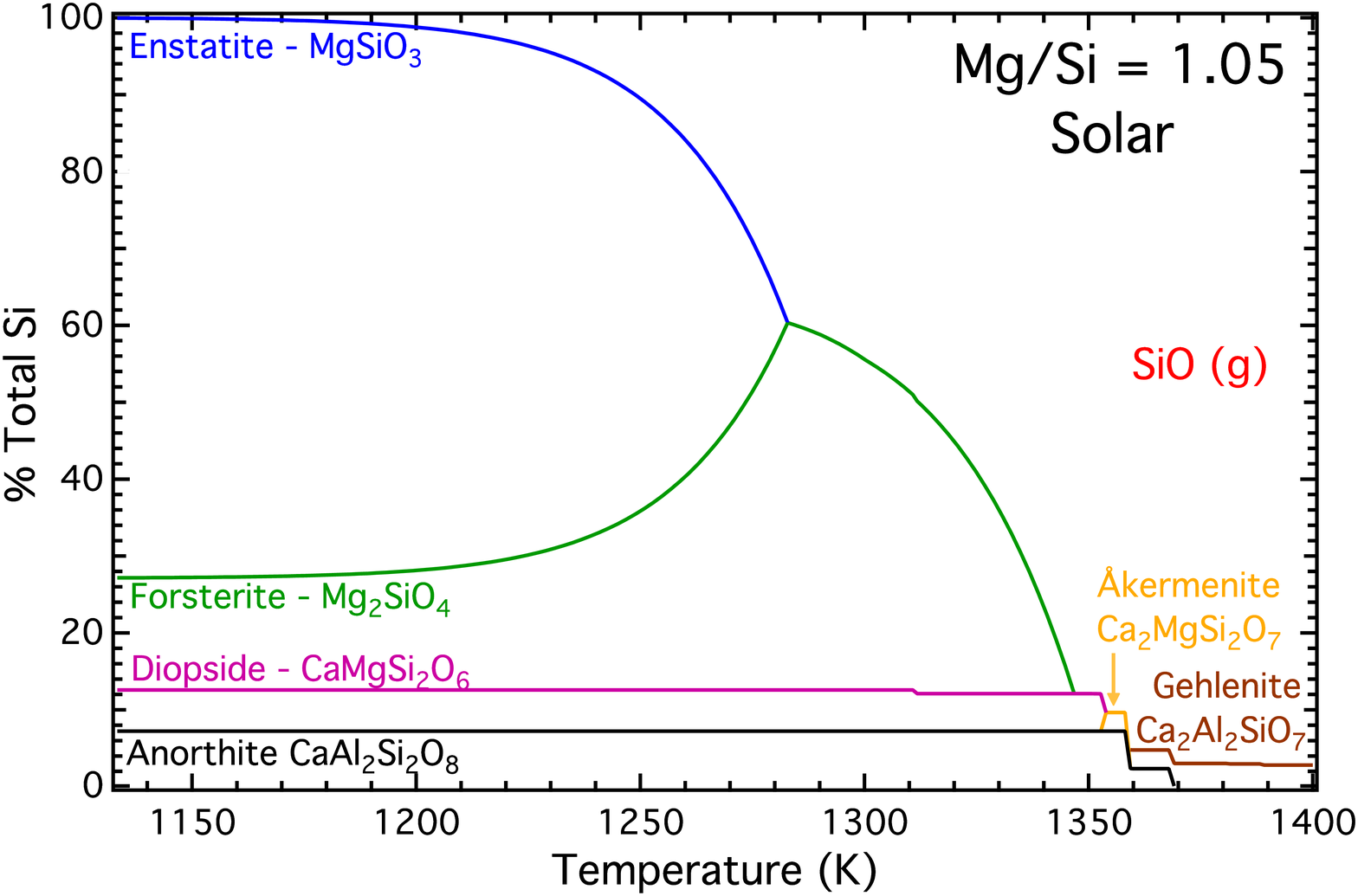}
\end{tabular}
\caption{Mg and Si phase diagrams for condensation sequence calculations adopting the Solar composition of \citet{Aspl05} as input. This figure represents the \textit{stable} phases at a given temperature and their relative proportions. For example, at 1283 K in the case of Mg, the stable equilibrium state is 91\% forsterite and 3\% diopside as solid phases and 6\% Mg gas. As temperature decreases, enstatite becomes stable and the relative proportion of forsterite decreases. This continues until at the point where all refractory elements are stable as solids (1133 K), forsterite accounts for the host of only $\sim$28\% of all Mg.} Both figures are on the same scale for comparison. Higher temperature condensed solids are shown as dashes.
\label{MgSiPhase_Solar}
\end{figure*}

\begin{figure*}
  \leavevmode
  \centering
   \epsfxsize=15cm
\includegraphics[width=15cm]{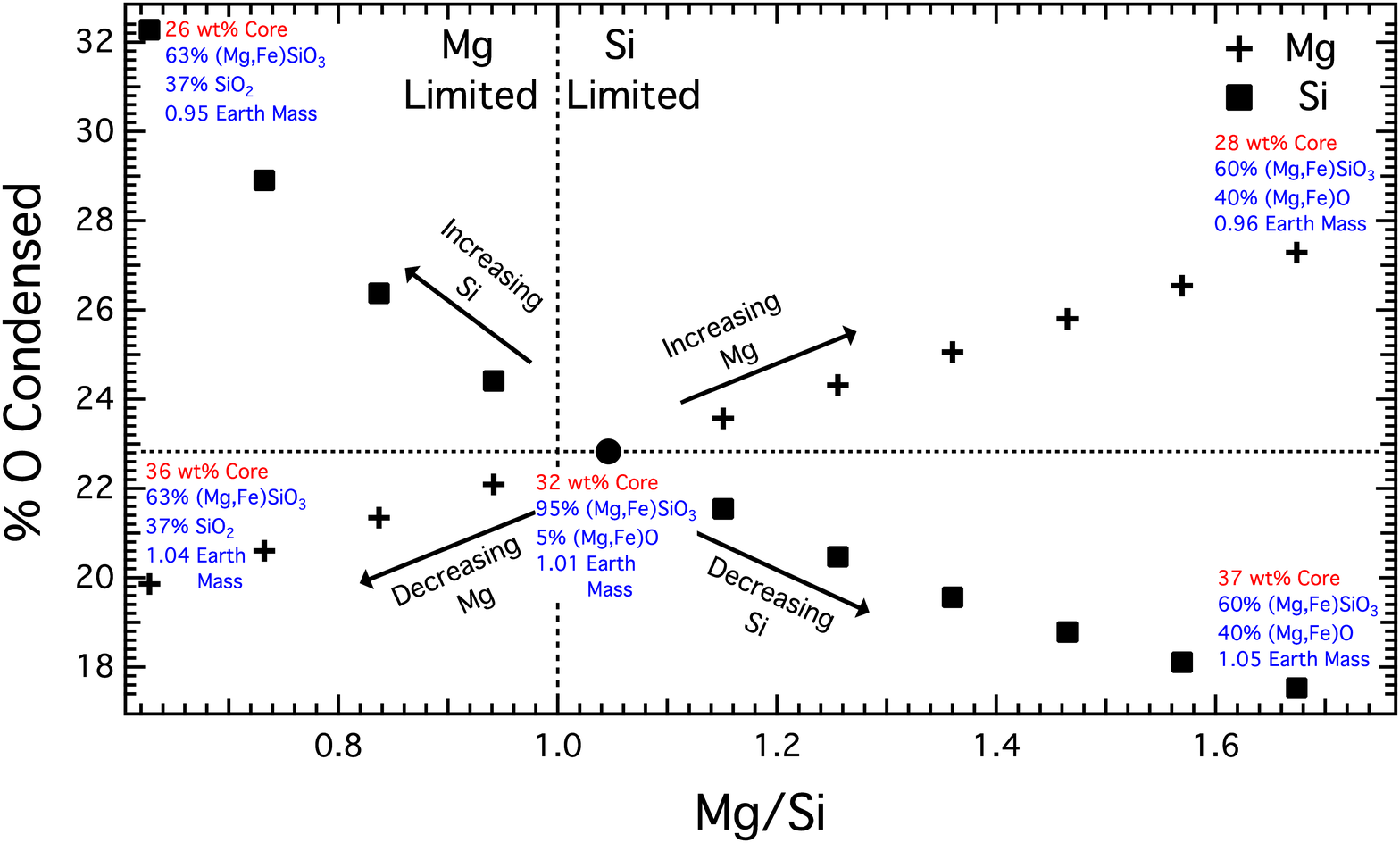}
\caption{The percentage of oxygen condensed in refractory phases as a function of Mg/Si for independent changes in both Mg (crosses) and Si (squares). The Sun \citep[circle;][]{Aspl05} is included for reference. Results of the stoichiometric determination of the core mass fraction (assuming only Fe-Ni alloy) and mantle mineralogy are appended for the Sun and each end-member. }
\label{ROvsMgSi}
\end{figure*}

\begin{figure*}
  \leavevmode
  \centering
   \epsfxsize=15cm
\begin{tabular}{cc}
\includegraphics[width=7cm]{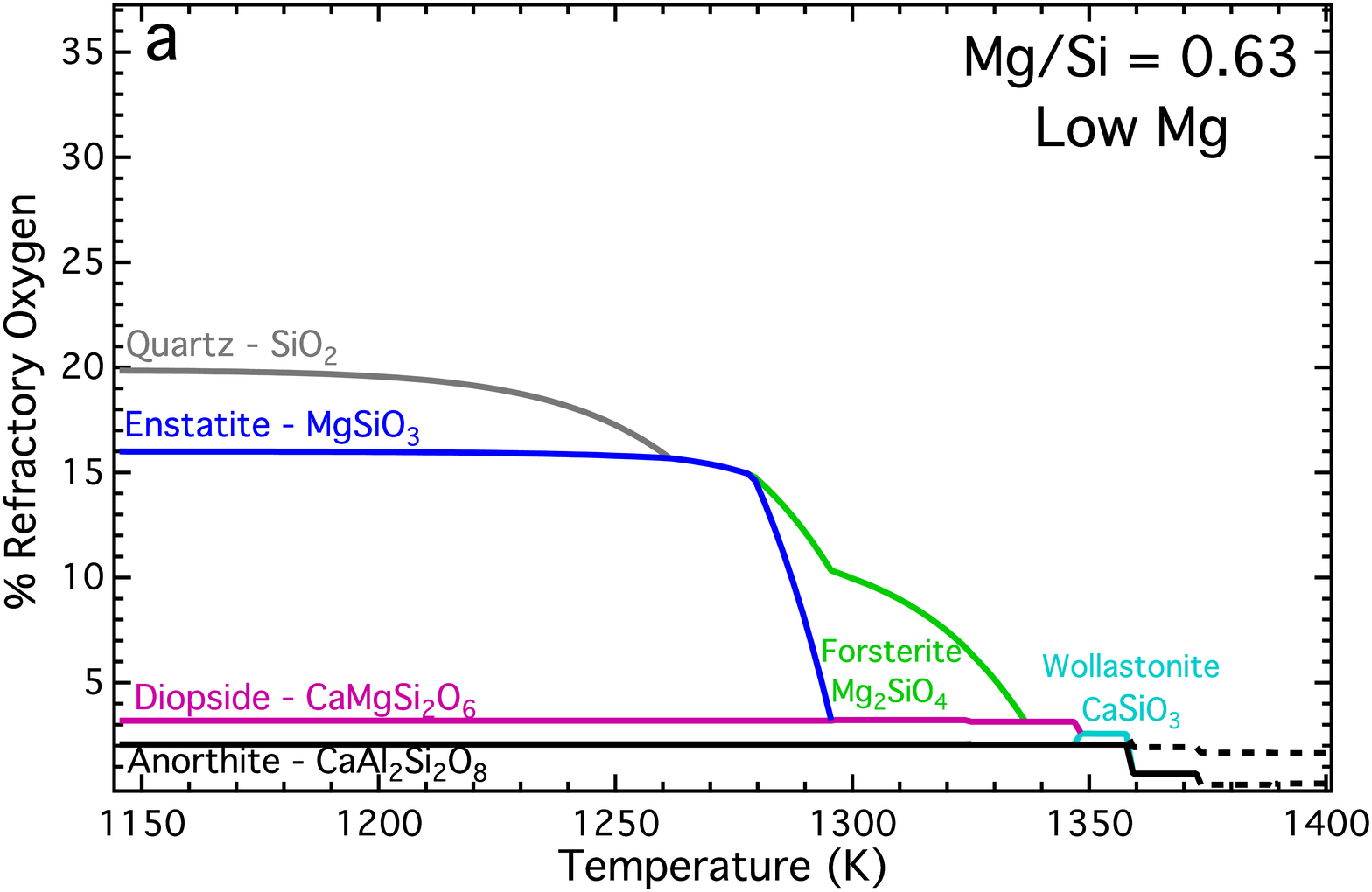} &\includegraphics[width=7cm]{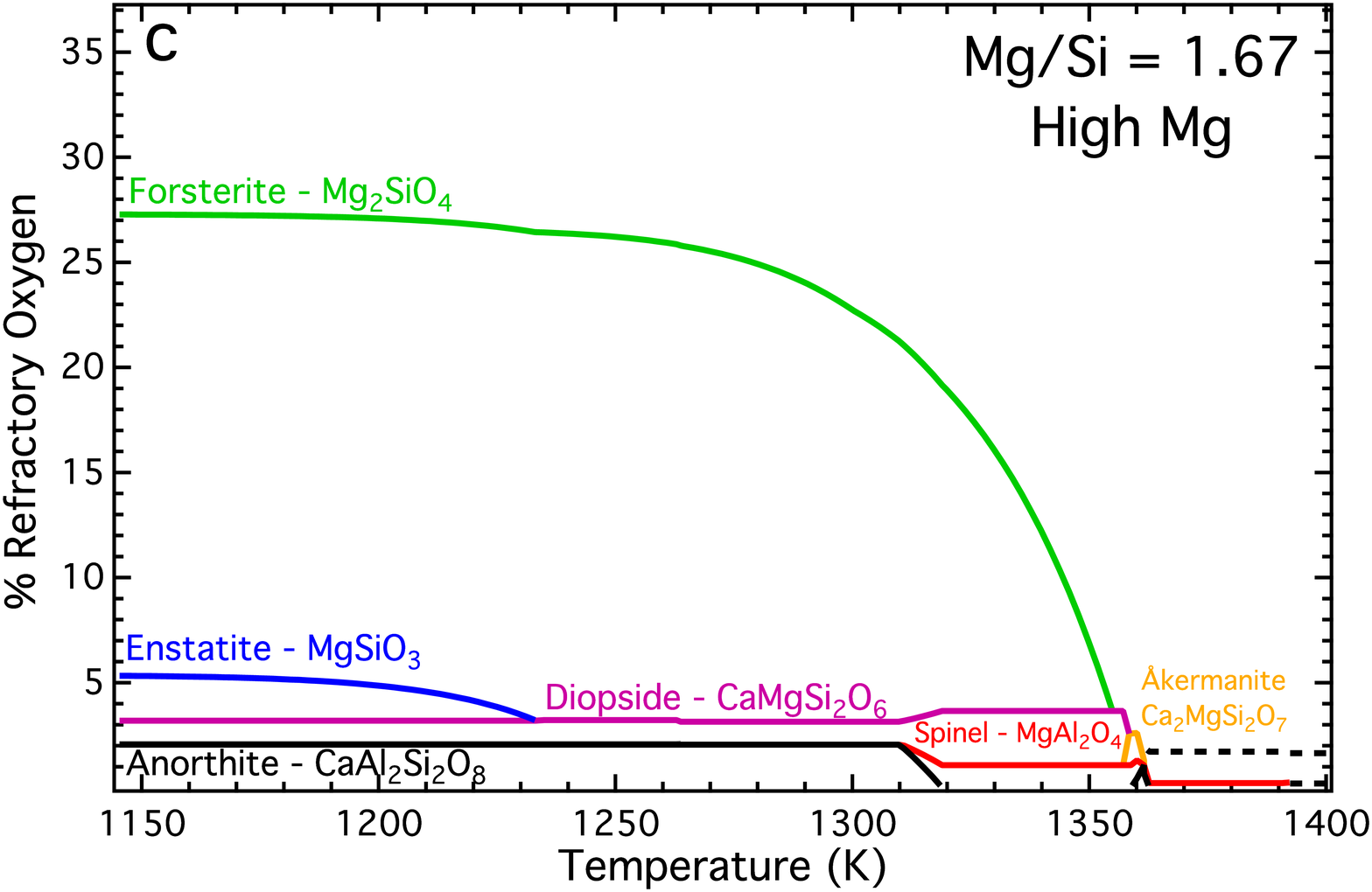} \\
\includegraphics[width=7cm]{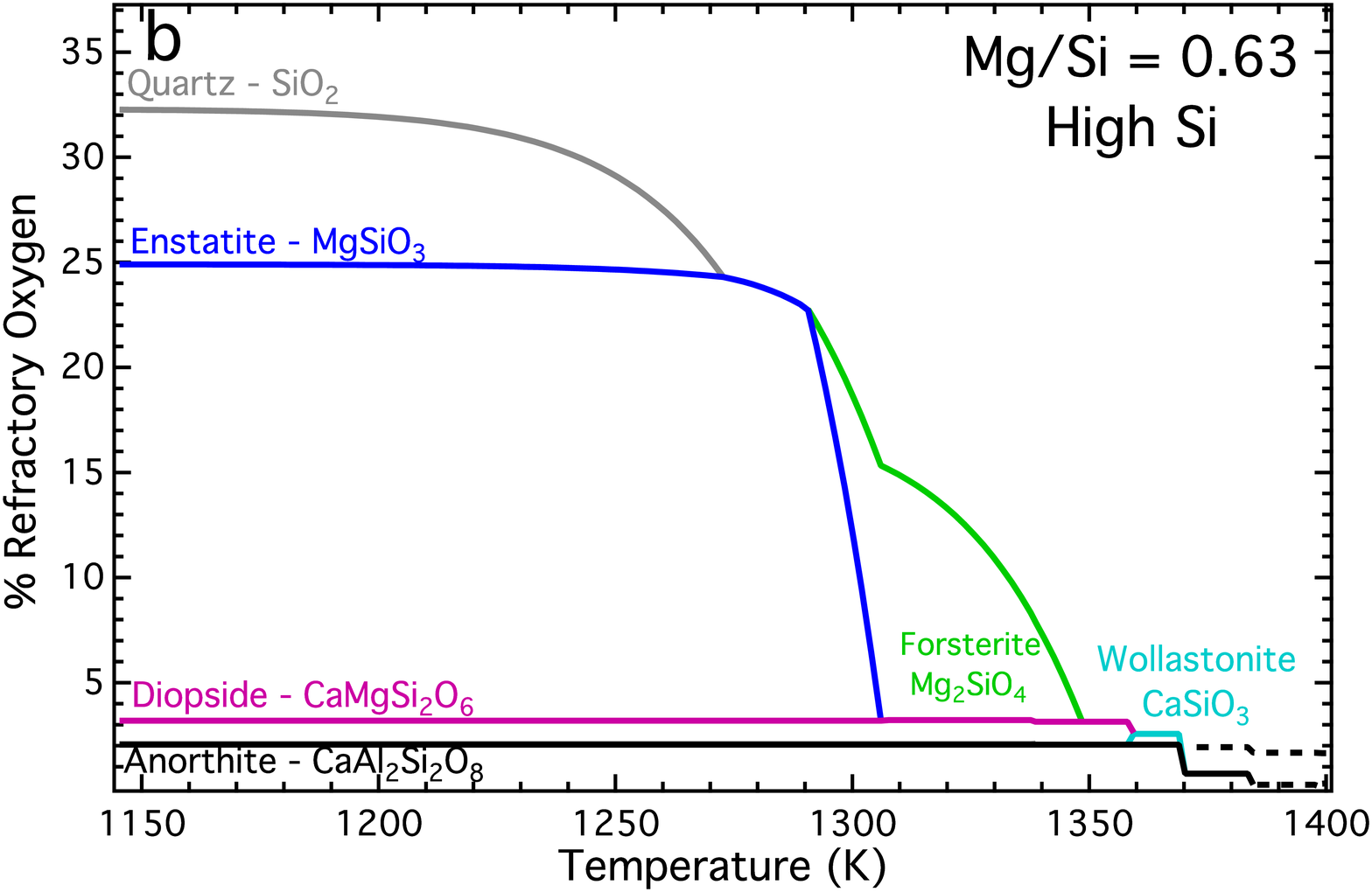} & \includegraphics[width=7cm]{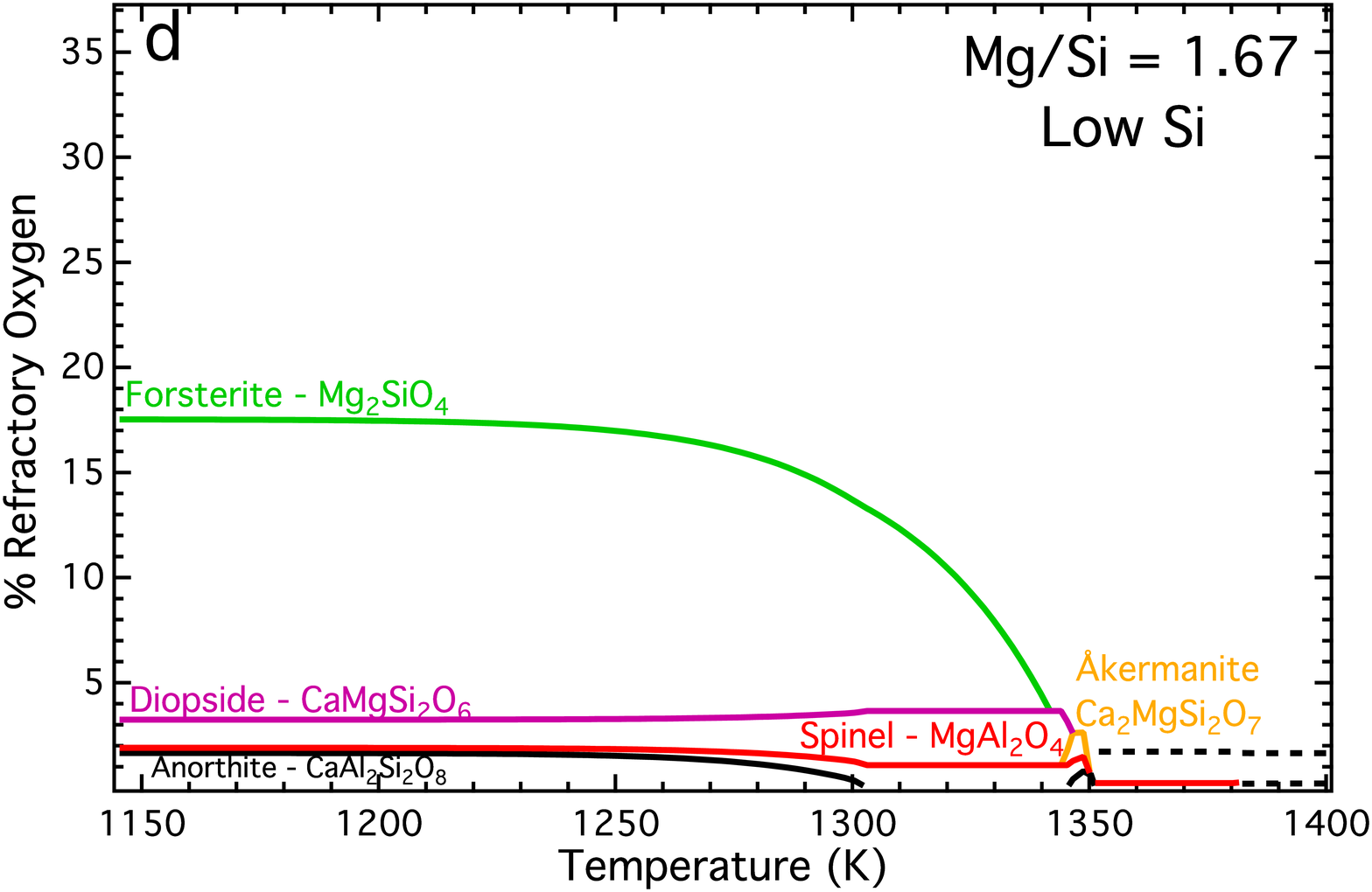}
\end{tabular}
\caption{Oxygen phase diagrams for high and low Mg/Si end-member calculations varying Mg and Si independently. All figures are on the same scale for comparison. Higher temperature condensed solids are shown as dashes and do not have a significant impact on the total \%RO.}
\label{ophase}
\end{figure*}

\indent The Solar model adopted by EG00 is that of \citet{Ande89} which has recently been revised by \citet{Grev93, Palm93, Grev96, Grev98,Grev02, Lodd03, Aspl05} and \citet{Aspl09}, the most recent of which includes a 3-dimensional, time-dependent hydrodynamical model of the solar atmosphere. This latter model, however, does not agree with that of the CI chondrites, which are often assumed to be indicative of composition of the Solar photosphere, particularly with respect to Mg. We therefore, adopt the next-most-recent Solar abundances of \citet{Aspl05} as our preferred Solar compositional model. \\
\indent The abundances of the major planet-building elements (Mg, Fe, Si) from the \citet{Aspl05} Solar model are within 10\% of the chondritic Earth model of \citet[][Table \ref{Molratio}]{McD03}. We find that when 100\% of refractory elements are stable in condensed solid phases, 22.8\% of Solar oxygen is condensed in refractory phases or 50.9\% of the total moles in the system, $\sim$5.5\% greater than \citet[][Table \ref{Molratio}]{McD03}. For this composition, enstatite (MgSiO$_{3}$) is the dominant host of O regardless of Solar model (Figure \ref{SolarO}). For \citet{Aspl05} (Figure \ref{SolarO}a), enstatite is responsible for 68\% of the total \%RO followed by forsterite (18\%), anorthite (9\%) and diopside (5\%). These phases are also the dominant hosts of both Mg and Si (Figure \ref{MgSiPhase_Solar}). While there is significant oxygen variability in Solar models, the condensation sequence for each of the models of \citet{Ande89,Lodd03,Aspl05} do not vary with respect to the relative proportions of the refractory elements or moles of oxygen condensed, but only in the fraction of oxygen condensed (13.7, 22.8, and 22.8\%, respectively, Figure \ref{SolarO}).\\
\indent To demonstrate the impact of variable stellar composition, we calculate the condensation sequences when varying Solar Mg/Si between 0.63 $\leq$ Mg/Si $\leq$ 1.67. We accomplish this in two ways: (1) varying Si between 159\% and 60\% of Solar (Mg/Si = 1.05) and (2) varying Mg between 63\% and 167\% of Solar, each while holding all other abundances constant. While neglecting associated variability in the other major elements is unrealistic, varying only Mg/Si represents a simplified case to illustrate the importance of how stellar composition can affect exoplanet interior composition and resulting core-mantle structure, as well as providing predictions of a terrestrial exoplanet's mass. \\
\indent We find that between 0.63 $\leq$ Mg/Si $\leq$ 1.67, in case (1), changes in Si results in a 14\% difference in the percent of refractory oxygen (18-32 \%RO) compared to case (2) in which \%RO varies by 8\% (20-28 \%RO) by exclusively changing Mg (Figure \ref{ROvsMgSi}). Furthermore, for constant Mg/Si, cases (1) and (2) result in different \%RO. As with the calculations of Solar composition, enstatite, forsterite, anorthite and diopside are the dominant host phases for O, with quartz becoming a major host at lower Mg/Si (Figure \ref{ophase}a, b). For a given change in Mg/Si though, we find that Si abundances more drastically affect \%RO compared to those in Mg. For example, when the Solar Si abundance is decreased by 60\%, \%RO increases by $\sim$10\% compared to Solar, whereas a similar increase in Mg only increases \%RO by $\sim$5\%. A similar result is found when Mg or Si is decreased. This behavior is due to the different stable oxidation states of each cation, Si$^{4+}$ and Mg$^{2+}$, thereby binding two (SiO$_2$) and one (MgO) condensed oxygen atoms. Therefore changes in the total silicon abundance will have a greater effect on \%RO.
\subsection{Discussion}
\indent The Earth's composition and differentiation into metal core and silicate crust and mantle are a reflection of the protoplanetary disk of Solar composition from which the Earth formed as well as any fractionation which occurred during planetary formation. As refractory elements, the dominant planetary cations Mg, Si, Fe, Al, and Ca are not expected to fractionate relative to each other considerably during planet formation. Combining the abundances of these elements with the oxygen budget predicted by ArCCoS allows us to estimate the relative mass proportion of oxidized mantle phases to core. For example, given the \citet{Aspl05} Solar composition and stoichiometric oxidation first of Mg, then Si, Al, Ca, Ti and finally Fe, 99\% of the total oxygen budget is consumed by Mg, Si, Ca, and Ti of which 95\% is due to Mg and Si alone (Table \ref{stoich}). The remaining oxygen oxidizes $\sim$14\% of the Fe, leaving the remainder in the reduced, metallic form. If all of this metallic iron and nickel segregate into the Earth's core, it accounts for 31.9 wt\% of the planet, or $\sim$99\% of the core's actual mass. \\
\indent This, however, neglects the presence of light elements (e.g. S, Si, O) in Earth's core. The presence of these light elements within the Earth's core is necessary to account for the seismically observed density difference between the core's mass and that of a pure Fe-Ni alloy \citep{Birch52,Jean79}, and is consistent with the observation of such light elements in some chondritic meteorites \citep{Weis06}. The Earth's core density can be constrained by its internal structure as well as major element ratios of the Earth assuming a density that is $\sim$94\% of pure-Fe \citep{Unte16}. Assuming Si and O are the dominant light elements present in the core and enter the core at a ratio of Si/O $\sim$3 \citep{Fisc15}, $\sim$7 mol\% of all Si must be present in the core to account for the 6\% mass deficit. Conserving the mass of the system, this incorporation of Si and O into the core reduces the mass of the core to 29.4 wt\%, or 91\% of the Earth core mass. \\
\begin{deluxetable}{lcc|lc}
\tablecolumns{5}
\tablecaption{Stoichiometric oxidation results for calculated condensation sequence of the Solar composition of \citet{Aspl05} for both a pure Fe/Ni core and an Fe/Ni core with 4.1 wt\% Si and 0.77 wt\% O$^{\ddagger}$ (\%RO = 22.8). Molar abundances are normalized so that there are only 100 moles of O available to oxidize material (\%RO*O = 100). \label{stoich}}
\tablehead{\colhead{} & \colhead{Abundance} & \colhead{\%O}&\multicolumn{2}{c}{Resulting planet} \\
\colhead{Oxide} & \colhead{(mol)} & \colhead{remaining}&\multicolumn{2}{c}{properties}}
\startdata
\cutinhead{Fe/Ni Core}
MgO&32.5&67.5&Core&32.0 mass\% \\
SiO$_{2}$&31.0&5.5&&\\
Al$_{2}$O$_{3}$&1.1&2.2&\multicolumn{2}{c}{\underline{Mantle}}\\
CaO&2.0&0.2&MgSiO$_{3}$&95.3 mol\% \\
Ti$_{3}$O$_{5}$&0.02&0.1&MgO&4.5 mol\% \\
FeO&0.1&0.0&FeO&0.2 mol\% \\
Core$^{\dagger}$&29.6&\nodata&Mg/Si&1.05\\
\cutinhead{Fe/Ni/Si/O Core}
MgO&32.7$^{*}$&67.3&Core&29.4 mass\% \\
SiO$_{2}$&29.1&9.2&&\\
Al$_{2}$O$_{3}$&1.13&5.8&\multicolumn{2}{c}{\underline{Mantle}}\\
CaO&2.0&3.8&MgSiO$_{3}$&80.0 mol\%\\
Ti$_{3}$O$_{5}$&0.02&3.7&MgO&10.0 mol\%\\
FeO&3.7&0.0&FeO&10.0 mol\%\\
Core$^{\dagger}$&28.0&\nodata&Mg/Si&1.12\\
\enddata
\tablerefs{\scriptsize $^{\dagger}$: Remaining after all O exhausted by formation of oxides. $^{\ddagger}$: Core contains 7.7 mol\% Si, 5.8 mol\% Ni and 2.6 mol\% O. $^{*}$: Normalized such that \%RO*O - O$_{core}$ = 100}
\end{deluxetable}
\indent As the balance of the oxidized material after core formation forms the silicate Earth, we can use this stoichiometric method to also estimate the relative proportions of mantle phases. Assuming the mantle mineralogy is that of iron-bearing bridgmanite ((Mg,Fe)SiO$_{3}$) as formed through the reaction of ferropericlase ((Mg,Fe)O) and silica (SiO$_{2}$) and all other elements (Ca, Al, Ti) are fractionated into the melt-extracted crust, we calculate a depleted mantle composition of $\sim$80\% bridgmanite and a remaining 10\% each of both periclase and w{\"u}stite (Table \ref{stoich}). This mantle Mg/Si, then, is $\sim$1.12, which is lower than the 1.25 of the chondritic Earth model \citep{McD95,McD03}. This latter value, is heavily weighted towards xenolith rock samples coming from Earth's upper-mantle. There is considerable debate as to whether the mantle is compositionally homogeneous, in which the Mg/Si of the upper-mantle may not be characteristic of the bulk mantle \citep{Mata07,Javo10}. These studies predict lower average mantle Mg/Si compared to the upper-mantle weighted, ``pyrolitic'' model.\\
\noindent While this simplistic stoichiometric model underpredicts the relative size of the Earth's core to mantle under realistic core compositional conditions, this underestimate is likely due to our predicted relative oxygen abundance being greater than the Earth model of \citet{McD03} (Table \ref{Molratio}). This overabundance of O causes more Fe to oxidize to FeO, thereby lowering the mass of the core. When Si is included in the core, this mass deficit is exacerbated due to the creation of fewer moles of mantle SiO$_{2}$. The oxygen that would be oxidized by Si then goes on to oxidize Fe instead, reducing the size of the core further. To lower, \%RO then, some fraction of the oxidized refractory elements must condense as reduced, rather than oxidized phases. A likely source of this reduction is the incorporation of light elements into metallic Fe during the condensation process. In the case of Si, any amount that is incorporated into Fe rather than oxidized phases such as enstatite or forsterite, will return 2 mol of O back to the gas phase, thus lowering \%RO while still allowing for the condensation of refractory elements. We are currently working to address these discrepancies by including solid-solution models within ArCCoS for the incorporation of light elements (e.g. S, Si) into condensing Fe-alloy, as well as FeO in olivine. It should be noted, however, that for 1 Earth-radius planet of Solar composition and an Earth-like core composition model has a planetary mass only 4\% smaller than the Earth's true mass and well within the current observational error for planetary mass \citep{Cott14,Unte16}. Thus we predict the bulk structure and mineralogy of the Earth within the current observational uncertainty and these more detailed calculations are beyond the scope of this paper. 
\subsection{Effects of Varying Mg/Si}
\begin{deluxetable}{lcc|c}
\tablecolumns{4}
\tablewidth{0pt}
\tablecaption{Comparison of the abundances of the major, planet-forming elements between \citet{McD03} and our calculation of percent refractory oxygen adopting the Solar model of \citet{Aspl05} \label{Molratio}}
\tablehead{\colhead{} & \colhead{\citet{Aspl05}} & \colhead{\citet{McD03}} & \colhead{}\\
\colhead{Element} & \colhead{mol \%} & \colhead{mol \%} & \colhead{$\%$ Difference}}
\startdata
Fe&13.8&15.2&-9.5\\
Ni&0.83&0.82&+3.4\\
Mg&16.5&16.8&-1.8\\
Si&15.8&15.2&+3.9\\
Al&1.1&1.5&-25.3\\
Ca&1.0&1.1&-11.2\\
O&50.9$^{\dagger}$&49.3&+3.4\\
Sum&99.93&99.92&\\
\enddata
\tablerefs{\scriptsize $^{\dagger}$: \%RO = 22.8}
\end{deluxetable}
\indent The consequences of variable Mg or Si on \%RO (Figure \ref{ROvsMgSi}) point to the importance of changes relative abundances of the planet-building cations on the stoichiometry and structure of the resulting planets. As the abundances of Si or Mg are changed, the relative change in \%RO depends on the stoichiometry of the condensates formed. In all cases, we find enstatite (MgSiO$_{3}$) or forsterite (Mg$_{2}$SiO$_{4}$) are the dominant host of O (Figure \ref{ROvsMgSi}). Therefore, the refractory oxygen abundance is limited to first order by the total amount of Si or Mg in the system. In the case of decreasing Mg or Si cases, the \%RO is a consequence of change in the total moles of these cations, thus limiting the total amount of enstatite and forsterite relative to Solar. In the case of increasing Mg, our calculations show the mineralogy favors a decrease in the relative proportion of enstatite (Mg/O = 3) to forsterite (Mg/O = 2, Figure \ref{ophase}c). When Si is increased (Figure \ref{ophase}b), however, forsterite (Si/O = 4) begins to transform into enstatite (Si/O = 3) and eventually quartz (SiO$_{2}$, Si/O = 2). It is this shift from Si/O = 4 to Si/O = 3 with increasing Si while Mg/O shifts from Mg/O = 3 to Mg/O = 2 with increasing Mg, that explains how changes in Si affect \%RO more so than the same change in Mg. Applying the same stoichiometric oxidation and structure models as with the Solar model, we find the mass for a 1 Earth radius planet given our calculated changes in core mass percentage to vary between 0.95 and 1.05 Earth masses (Figure \ref{ROvsMgSi}). While small, this variation provides a potential observational test of this model as the uncertainty in mass measurements improve. \\
\indent These calculations reveal that while at a given Mg/Si, the mantle mineralogy is the same, however, the relative mass of the core to the rest of the planet can vary by $\sim$10\% depending on whether Mg or Si is varied. As alpha elements, both Mg and Si are thought to scale together in abundance, with any variations away from this trend being potentially due to their exact nucleosynthetic origin \citep{Adi15}. We demonstrate here, that even differences from Solar as small as of $\pm0.2$ dex in either a host star's Mg or Si abundance can impact the bulk structure and mineralogy of an orbiting terrestrial planet. These results point not only to the importance of a star's bulk Mg/Si in determining the a planet's potential mineralogy and structure, but also the absolute abundances of Mg, Si (and Fe). That is, \textit{no single ratio of elements} is sufficient to characterize the chemical state of a planet and observational uncertainties must be low in order to make any substantial comparisons between stellar/planetary systems.
\section{Conclusion}
\indent We present here the Python-based, open-source software package, Arbitrary Composition Condensation Sequence Calculator (ArCCoS). It is designed for calculating the stability of solid phases in equilibrium with gas with a wide variety of applications including determining the composition of the first solids in an exoplanetary system during formation, dust condensation in molecular clouds and the interstellar medium and any other application where solid/gas phase equilibria are necessary. To date though, all software to calculate these condensation sequences published to date is commercially available or closed-source. Here, we utilize ArCCoS, combined with simple stoichiometry, can reproduce the Earth's bulk structure and total mass to within observational uncertainty. Furthermore, we show here that changes in stellar Mg and Si abundances, and thus bulk Mg/Si, affect the relative core size of exoplanets and their mantle mineralogy. These changes provide observable differences in total planetary mass for a given stellar composition. This model assumes the composition at 100\% refractory element condensation is representative of the ``average'' terrestrial planet in a system. We are working to address the more complicated question of planets forming within specific radial ``feeding-zones'' where local compositional and oxygen fugacity gradients may exist, may be rather than adopting this average composition and the effect this may have within multi-planet systems as well as incorporated more sophisticated solid-solution models into ArCCoS. Such improved models are likely required address the modeled discrepancy of Venus mass-radius models not being well fit by an Earth or Solar composition \citep{Ring76,Unte16}.
\acknowledgements 
This work is supported by NSF CAREER grant EAR-60023026 to WRP. 
\bibliographystyle{apj}

\end{document}